\documentclass[11pt]{article}
\input epsf

\title{\bf Nucleus-nucleus cross-sections and long-range correlations
with a local supercritical pomeron}
\author{M.A.Braun\\
Dep. of High Energy physics,
 University of S.Petersburg,\\
198504 S.Petersburg, Russia  and \\
Dep. of Particle Physics, University of Santiago de Compostela,\\
15782, Santiago de Compostela, Spain }
\def\beq{\begin{equation}}
\def\eeq{\end{equation}}
\def\noi{\noindent}
\def\phid{\phi^{\dagger}}
\def\tchi{\tilde{\chi}}
\def\tg{\tilde{G}}
\begin{document}
\maketitle
\medskip
\noi{\bf Abstract.}

Nucleus-nucleus scattering is studied in the local Reggeon
Field Theory in the quasi-classical approximation
with non-eikonal boundary conditions corresponding
to the Glauber picture at low energies. Comparison with the commonly used
eikonal boundary conditions shows that the new conditions  make
both the action and nucleus-nucleus total cross-sections lower by
$3\div 5$ \%. They also substantially change the behaviour of the
solutions of the equations of motion at low energies. Using expressions 
for the double inclusive cross-sections derived earlier in the Reggeon 
Field Theory ~\cite{CM} long-range rapidity correlations are calculated
for the RHIC and LHC  energies.

\section{Introduction}
Recent and forthcoming collider experiments attract much attention to the
description of nucleus-nucleus collisions. The corresponding amplitudes are
very complicated already in the pure Glauber approach and become still more
difficult to calculate in the pomeron approach with the triple pomeron
interaction. In a recent publication the nucleus-nucleus amplitude
was calculated within the tree approximation to the effective field theory
of interacting perturbative QCD (BFKL) pomerons ~\cite{BB} taking
into account non-eikonal boundary conditions, which are required to ensure
the correct structure of the amplitude at low energies. It was found that
in contrast to the situation with the eikonal boundary conditions, a whole set 
of different
solutions follow from the relevant classical equations of motion. Both the
mathematical reason  and physical significance of this phenomenon are unclear.
Is the higly non-linear form of the boundary conditions or rather
the interplay between them and the structure of the BFKL pomeron responsible?
A possible way to study this problem is to turn to a much simpler case
of the old  local pomeron described by the Local Reggeon Field Theory
(LRFT) with the triple pomeron interaction, studied long ago with eikonal
boundary conditions analytically
in a series of elegant papers (see ~\cite{ACJ} and references therein). Apart 
from
the pure theoretical interest, such a study may have a direct relation
to the experimental situation, to the extent in which soft high-energy
collisions can be  described in the framework of LRFT.

Our results indicate that introduction of the improved non-eikonal boundary
conditions into LRFT does not lead to dramatic changes which are observed
in the QCD perturbative approach. True, it lowers the value of the action
and also of the resulting cross-sections, as in the latter approach. However
this effect is not at all large and, most significant, solutions of the
classical equation do not infinitely multiply. So it is the interplay of the 
boundary
conditions and inner structure of the pomeron, which leads to appearance of
a family of solutions their number growing  with rapidity.
In fact behaviour of the solutions with rapidity is opposite: with its 
growth
the number of solutions becomes smaller.
From the experimental point of view, the predicted
nucleus-nucleus cross-sections are much smaller than in the optical
approximation, in accordance with the conclusions in ~\cite{ACJ}.
The correction introduced by the non-eikonal boundary conditions is
however relatively small.

The found  cross-sections allow us to study the long-range
rapidity correlations in particle production in nucleus-nucleus collisions,
which have recently attracted much attention.
In  ~\cite{ALP} qualitative predictions
for the correlation coefficient $\sigma_{FB}$ of the forward-backward
correlations have been presented in the framework of the the Color Glass
Condensate model. It was argued that this coefficient should fall with
the rapidity distance $\Delta y$
between the two windows and with the impact impact parameter and grow with 
the overall energy. However
no concrete numerical values were obtained. In this paper we remind that
in the framework of LRFT fully quantitative predictions for the correlations
can be obtained. Of course this formalism does not allow to study the $p_t$
dependence of the inclusive spectra, which are to be treated either in the
framework of the QCD (for the hard part of the spectra) or in models
specially oriented to the soft part of the spectrum, such as colour string models.
However for the long range correlations
in rapidity of  the spectra integrated over the transverse momenta 
LRFT seems to be fully adequate. So much the more that the highly non-trivial
expression for the double inclusive cross-section in nucleus-nucleus
collisions, necessary for the correlations, has up to the present
been derived only in LRFT ~\cite{CM}.

In  ~\cite{CM} also the asymptotic expressions for the single and double
inclusive cross-sections were found. However they refer to the case when
all rapidity intervals, including the rapidity distance $\Delta y$, are 
very large, which  does not correspond to the
experimental setup. So in this paper, without introducing substantial new
ingredients, we calculate the correlation coefficient $\sigma_{FB}$ from
the old formuals and the found cross-sections
for the present and future experimental situations.

The supercritical pomeron LRFT contains essentially three paramenters:
the pomeron intercept $\alpha(0)=1+\mu$, its slope $\alpha'(0)$ and
three-pomeron coupling $\lambda$. The first and the last are
more or less known. As for the slope, we take it equal to zero, since,
on the one hand, all that is definitely known about it is that it is small
and, on the other hand, its influence in heavy nuclear collisions should
be expected only at energies much higher than the present and
forthcoming ones. Last but not least, the problem
with $\alpha'(0)=0$ drastically simplifies and in fact allows to obtain
analytical solutions for the inclusive spectra at a given point in
the nuclear overlap area.

In fact the LRFT does not allow to calculate the correlation coefficient
without any additional parameters. The contribution from the
short-range correlations
corresponding to the simulteneous production of the two observed particles
from the pomeron is to be added to the rest part with an unknown weight $C$.
This weight is the only new parameter which we introduce. Its good quality
is that it is universal: it should not
depend on any of the external variables coming from the experimental setup:
$\Delta y$, overall rapidity $Y$, impact parametr $B$ atomic numbers of the
collising nuclei etc. Its value can be inferred by comparison to the values
of $\sigma_{FB}$ measured at RHIC for Au-Au collisions.

Our results confirm the overall behaviour of ther correlation coefficient
found in ~\cite{ALP}: it grows with the overall energy and falls with
$\Delta y$ and $B$.

The paper is organized as follows. In the next section we briefly recapitulate 
the basics of the LRFT. Section 3 is devoted to discussion of the boundary 
conditions for the equations of motion for the pomeronic fields.
In Section 3. we present our numerical results for the total 
nucleus-nucleus cross-sections. 
In Section 4. they are used for calculation of the 
correlation coefficieint of the long-range rapidity correlations.
Here also the necessary formulas for the single and double inclusive 
cross-sections are reproduced following ~\cite{CM}.
The last section draws some conclusions


\subsection{Action of the LRFT and classical equations of motion}
The $S$-matrix for  nucleus-nucleus collisions at impact parameter
$B$ and rapidity $Y$ in the supercritical local pomeron model with
a three-pomeron interaction is given by
 \beq S(Y,B)=e^{-A_{min}},
\eeq
where $A$ is the classical action
\beq
 A=\int dyd^2b{\cal L}
  \eeq
 in the presence of external sources, which only act at rapidities of
the participant nuclei $y=0$ and $y=Y$.
 The Lagrangian is given by the
sum of three terms
\beq
 {\cal L}={\cal L}_0+{\cal L}_I+{\cal L}_E,
\eeq
 which represent free, interaction and external parts. The
first two are given by
\beq
 {\cal L}_0=
\frac{1}{2}(\phid\dot{\phi}-\dot{\phid}\phi)+\mu \phid\phi-
\alpha'\nabla\phid\nabla\phi,
 \eeq 
\beq 
{\cal L}_I= - \lambda\phid\phi (\phid+\phi).
\eeq
 The form of the external part ${\cal
L}_E(\phi,\phid)$ will be discussed in the next section. In these
expressions $\phi(y,b)$ and $\phid(y,b)$ are two real fields, which
describe in- and out-going pomerons with the intercept
$\alpha(0)=1+\mu$, slope $\alpha'$ and  three-pomeron coupling
$-\lambda$. $\dot{\phi}$ denotes the derivative in $y$, $\nabla$
refers to $b$. Due to smallness of the slope, following
~\cite{ACJ} we neglect the kinetic energy of the pomeron and study
the simple case $\alpha'=0$.

The classical equations of motion are then (at point $(y,b)$)
\beq
0=\delta A/\delta \phid=\dot{\phi}+\mu\phi-\lambda\phi^2-2\lambda
\phid\phi
\label{phieq}
\eeq
and
\beq
0=\delta A/\delta \phi=-\dot{\phid}+\mu\phid-\lambda{\phid}^2-2\lambda
\phid\phi
\label{phideq},
\eeq
with the boundary conditions
\beq
\phi(0,b)\delta(y)=d{\cal L}_E/d\phid(y,b),\ \
\phid(Y,b)\delta(y-Y)=d{\cal L}_E/d\phi(y,b),
\eeq
which follow from the conditions
\beq
\phi(y,b)=0\ \ {\rm at}\ \ y<0,\ \
\phid(y,b)=0\ \ {\rm at}\ \ y>Y.
\eeq

Using the equations of motion one can find that parts of the classical action
satisfy the relation
\beq
A_0=-\frac{3}{2}A_I+\frac{1}{2}\int d^2b\Big(\phid(0,b)\phi(0,b)+
\phid(Y,b)\phi(Y,b)\Big),
\eeq
which allows to exclude one part (say $A_0$) and write the classical action
as
\beq
A=-\frac{1}{2}A_I+A_E+\frac{1}{2}\int d^2b\Big(\phid(0,b)\phi(0,b)+
\phid(Y,b)\phi(Y,b)\Big).
\eeq

\section{External sources}
Each nucleon from the participant nuclei
may emit any number of pomerons, so that the
source term for nucleons in nucleus A is
\beq
AT_A(b)\delta(y)\Big(e^{-g\phid(0,b)}-1\Big)
\eeq
and the source term for nucleons in nucleus B is
\beq
BT_B(B-b)\delta(y-Y)\Big(e^{-g\phi(Y,b)}-1\Big).
\eeq
Here $T_{A(B)}$ are the standard nuclear profile functions,
normalized to unity, $g$ is the nucleon-pomeron coupling.
Accordingly $S$ matrix at a fixed overall impact parameter $B$ will be
given by
\[
S(Y,b)=\int D\phi D\phid e^{-{ A}_0-{ A}_I}
\sum_{n_A,n_B}C_A^{n_A}C_B^{n_B}\]\beq\int\prod_{i=1}^{n_A}
\prod_{j=1}^{n_B}d^2b_{i}d^2b'_{j}T_A(b_i)T_B(B-b'_j)
\Big(e^{-g\phid(0,b_i)}-1\Big)\Big(e^{-g\phi(Y,b'_j)}-1\Big),
\eeq
where actions ${ A}_0$ and ${ A}_I$ are the free and interaction
part of the action introduced in the preceding section.
Summations over $n_A$ and $n_B$ lead to the final formula
\beq
S(Y,b)=
\int D\phi D\phid e^{-{ A}_0-{ A}_I-{ A}_E},
\label{swinb}
\eeq
where the external action ${ A}_E$ is
\beq
{ A}_E=-A\ln\int d^2b'T_A(b')e^{-g\phid(0,b')}-
B\ln\int d^2b'T_B(B-b')e^{-g\phid(Y,b')}.
\label{winbterm}
\eeq

This form of the external sources leads to the boundary conditions
\[
\phi(0,b)=AgT_A(b)\frac{e^{-g\phid(0,b)}}{\int 
d^2b'T_A(b')e^{-g\phid(0,b')}},
\]\beq
\phid(Y,b)=AgT_B(B-b)
\frac{e^{-g\phi(Y,b)}}{\int d^2b'T_B(B-b')e^{-g\phi(Y,b')}}.
\label{nlbound}
\eeq
As one observes, these boundary conditions are not only non-linear but
they also mix values of the fields at $y=0$ and $y=Y$. So they do not
represent initial conditions for  evolution of the fields in rapidity
but rather relate initial and fully evolved values of the fields.

One also observes that the complexity of these non-eikonal boundary
conditions comes exclusively from the non-trivial $B$ and $b$
dependence of the nuclear distributions
$T_A(b)$ and $T_B(B-b)$. Indeed in the (fictitious) case of  constant
profile functions, for central collisions $B=0$, both $\phi$ and $\phid$
are constant in the common nuclear area and zero outside. Then conditions
(\ref{nlbound}) become equivalent to the standard eikonal
boundary conditions
\beq
\phi(0,b)=AgT_A(b),
\ \
\phid(Y,b)=AgT_B(B-b).
\label{eibound}
\eeq

\section{Solution}
\subsection{Technique}
As in our previous studies (~\cite{bra1,BB}) we searched for a solution
of the classical equations of motion with given boundary conditions by
iterations. We applied three different procedures.
Most generally, at the first step we constructed  
$\phi(y,b)$ by evolution in $y$
of Eq. (\ref{phieq}) with  $\phid(y,b)=0$ starting from
its initial condition at $y=0$ obtained from (\ref{nlbound}) with
$\phid(Y,b)=0$. At the next step we evolved $\phid(y,b)$ back according to
Eq. (\ref{phideq}) with already found values of $\phi(y,b)$ starting from its
value at $y=Y$ obtained from (\ref{nlbound}) with the found value of
$\phi(Y,b)$. Then this procedure is repeated. At each step
both the equation of motion and initial values for $\phi(y,b)$ are
taken with $\phid(y,b)$ found in the previous iteration and vice versa.
This procedure allows to find solutions which for identical nuclei
($A=B$) are not symmetric
in the interchange of the projectile and target, that is do not
satisfy the relation
\beq
\phi(b,y)=\phid(B-b,Y-y).
\label{sym}
\eeq
To specifically find such symmetric solutions we changed the procedure
either by evolving both $\phi$ and $\phid$ simulatenously
or simply
substituting evolution of $\phid(y,b)$ by directly expressing it
according to (\ref{sym}). Finally we repeated our calculations for
the eikonal boundary conditions  (\ref{eibound}).

\subsection{Results}
In accordance with the experimental data we have taken
\beq
\mu=0.08,\ \ g_N=8.4/\sqrt{2}\  {\rm GeV}^{-1}, \ \ 
\lambda=\frac{\sqrt{2}}{25}g_N=
0.48 {\rm GeV}^{-1}.
\label{par}
\eeq
(see Appendix).
For the nuclear profile functions we have taken those which are generated
by the Woods-Saxon nuclear density.
We considered collisions of identical nuclei with $A=64$ and 197.

Our first series of results refers to  values of the classical action at $B=0$
related to the $S$ matrix for central collisions as
\beq
S(Y,B=0)=e^{-A(Y,B=0)}.
\eeq
In Figs. 1 and 2 we show these values obtained for Au-Au collisons
with the non-eikonal
boundary conditions (\ref{nlbound}), eikonal boundary conditions
(\ref{eibound}) and in the optical approximation when the action is
\[
g^2AB\int d^2bT_A(b)T_B(B-b).
\]
\begin{figure}
\epsfxsize 4in
\centerline{\epsfbox{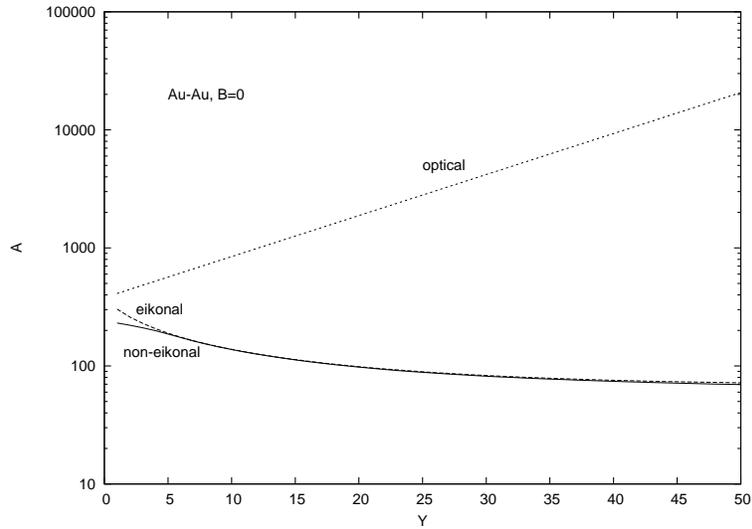}}
\caption{Minimal action obtained with the non-eikonal and eikonal
boundary conditions as well as in the optical approximation as
a function of rapidity for central Au-Au collisions}
\label{fig1}
\end{figure}
\begin{figure}
\epsfxsize 4in
\centerline{\epsfbox{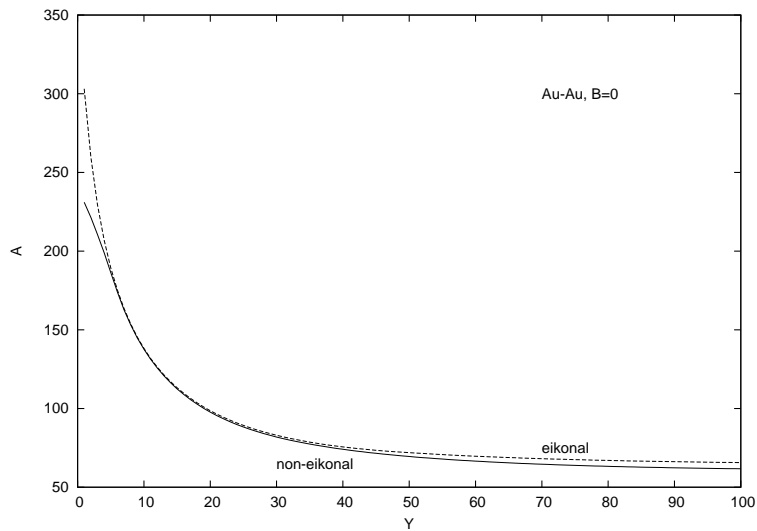}}
\caption{Minimal action obtained with the non-eikonal and eikonal
boundary conditions  in the extended interval of rapidities
for central Au-Au collisions.}
\label{fig2}
\end{figure}
As mentioned,
in contrast to the case of BFKL pomerons ~\cite{BB}, in this local
theory and with our choice of the couplings
the equations of motion with 
the non-eikonal boundary conditions
generate a finite number of solutions, different at rapidities lower
or larger than a certain critical one, which is numerically found to be
$Y_c\simeq 6$ . At $Y<Y_c$  one gets a pair of solutions, which are
not symmetric under the
interchange of projectile and target nuclei. 
At $Y>Y_c$ this pair merges into
a single symmetric solution. 
In Fig. 3 we show
values of the fully evolved fields $\phi$ and $\phid$ at the center ($b=0$).
As one observes, in the rapidity range $0<Y<6$ 
their values are very different but become equal at larger $Y$. 
\begin{figure}
\epsfxsize 4in
\centerline{\epsfbox{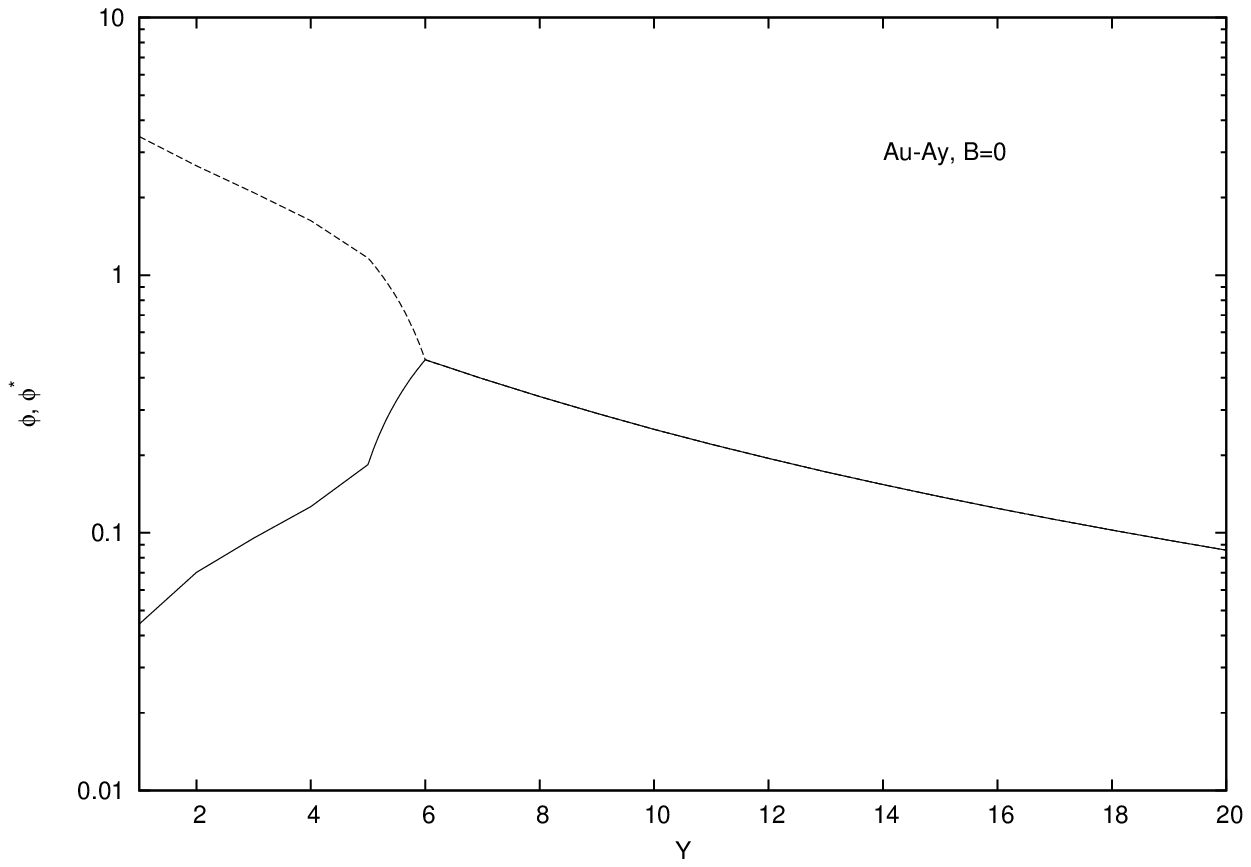}}
\caption{Values of the fully evolved fields  $\phi$ and $\phid$ at the center ($b=0$).
Below $Y=Y_c=6$ the upper-and lowermost curves correspond to
the asymmetric pair of solutions.} 
\label{fig3}
\end{figure}

Under the eikonal boundary eikonal conditions the situation repeats the
one found in \cite{BM}. At rapidities lower than $\sim 49$  a unique
symmetric solution is found. At higher rapidities from this
symmetric solution splits an asymmetric one, which realizes
the minimum of the action. Both solutions coexist up to $Y=82$ after which
our procedures generate only a pair of asymmetric solutions.

As expected, with both types of  boundary conditions the action
is found to be much smaller than in the naive optical approximation.
Moreover it slowly falls with rapidity.
whereas the optical action naturally
grows as the pomeron itself.
Going to rapidities as high as 1000 we found that for both types
of the boundary conditions the action freezes at values oly a
few percent smaller than shown in Fig. 2 at $Y=100$.
The difference introduced by the new
non-eikonal boundary conditions is quite significant only at
comparatively small
rapidities ($\sim 30$ \%) and in the high-energy limit  goes down 
to $\sim 6.5$\%.

At $B>0$ the situation with the non-eikonal boundary conditions is found
to be the following. Until $B$ is greater than $B_c=8$ fm
there exists a single symmetric solution, depending on the angle between 
$\bf B$ and $\bf b$
and realizing the minimum of the action,
and a pair of asymmetric solutions  As $B\to 0$ the symmetric solution
passes into the  pair of
asymmetric solutions. At $B>B_c$ only a single symmetric solution exists.
With the eikonal boundary conditions the problem is purely local, so that the
behavior of solutions is qualitatively the same as for $B=0$.

In Fig.4 we show values of the action for central Cu-Cu collisions.
Both with the non-eikonal and  eikonal boundary conditions the action
turns out to be proportional to $A^{2/3}$ as opposed
to $A^{4/3}$  naively expected from the optical approximation.
This dependence fully agrees with asymptotical estimates
in ~\cite{ACJ}. The structure of the solutions repeats that for
Au-Au collisions with the same $Y_c= 6$ and $B_c= 3.5$ fm.
\begin{figure}
\epsfxsize 4 in
\centerline{\epsfbox{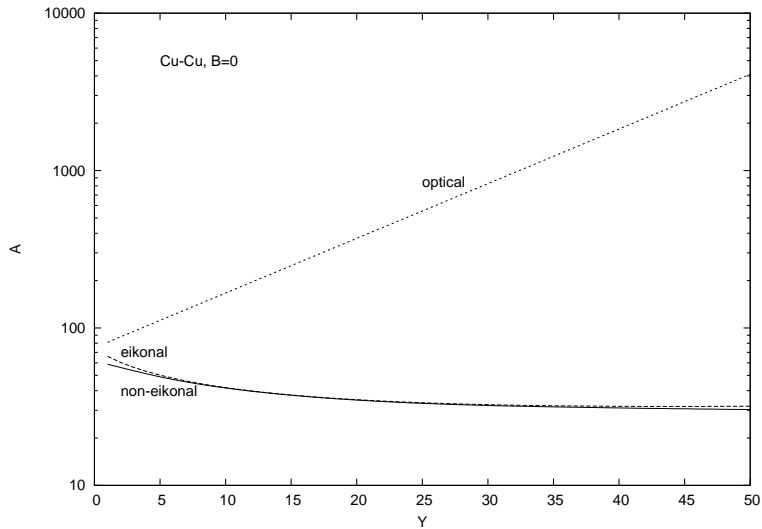}}
\caption{Same as Fig. \ref{fig1} for central Cu-Cu collisions}
\label{fig4}
\end{figure}

The total inelastic nucleus-nucleus cross-sections are given by an 
integral over
impact parameter $B$:
\beq
\sigma_{AB}^{in}=\int d^2B\Big(1-S^2(Y,B)\Big).
\eeq
Here, in principle, we have to take into account a possible
existence of several solutions of the classical equation of motion,
which contribute to the $S$-matrix
with different prefactors, determined by quantum corrections.
In the purely local case we can use the result of  ~\cite{ciaf}.
For the eikonal boundary conditions  in was found there that above the 
critical rapidity $Y_c$,  when from
a single symmetric solution splits a pair of asymmetric ones which give 
the minimum of the action, the $S$ matrix was given by
the sum of  contributions from the two asymmetric solutions minus
the one from the symmetric solution. This guarantees continuity of the 
$S$-matrix
at $Y=Y_c$. However this conclusion does not apply to the case of our 
non-eikonal boundary conditions, when both below and above the critical 
rapidity there may exist
several solutions.

Fortunately for the calculation of the total cross-section this problem
is of no relevance. In fact values for the action below the critical
impact parameter $B_c$ are very large both for Au-Au ($A>15.$) and Cu-Cu 
($A>10.$)
collisions. So the contribution to the $S$-matrix from these $B$ is quite 
small and
can safely be neglected altogether.  At $B>B_c$, on the other hand, the 
only contribution
comes from the single symmetric solution, so that the problem of relative 
weights does not arise.

The cross-sections calculated in this way are shown in Fig. 5 and 6 for 
Au-Au and Cu-Cu collisions
respectively  As one observes, the cross-sections calculated with the
corrected non-eikonal boundary conditions, eikonal boundary conditions and
in the optical approximation all rise with rapidities albeit with a different
rate. This rise is obviously due to the contribution from the peripheral
collisions when the action becomes small. Physically this corresponds to
the growth of the effective nuclear radius  with energy. In the naive
optical approximation the cross-sections grow linearly with $Y$. With
both the eikonal and non- eikonal boundary conditions this rise is weaker.
Absolute values of the cross-sections are correspondingly the largest
in the optical approximation and smallest with the non-eikonal boundary
conditions, the difference between them growing with rapidity.
\begin{figure}
\epsfxsize 4 in
\centerline{\epsfbox{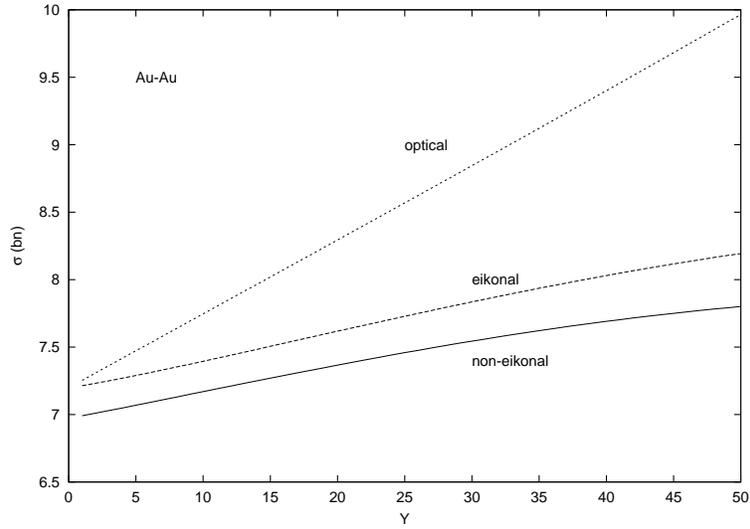}}
\caption{Total inelastic Au-Au cross-sections obtained with the 
non-eikonal 
and eikonal
boundary conditions as well as in the optical approximation as
a function of rapidity}
\label{fig5}
\end{figure}
\begin{figure}
\epsfxsize 4 in
\centerline{\epsfbox{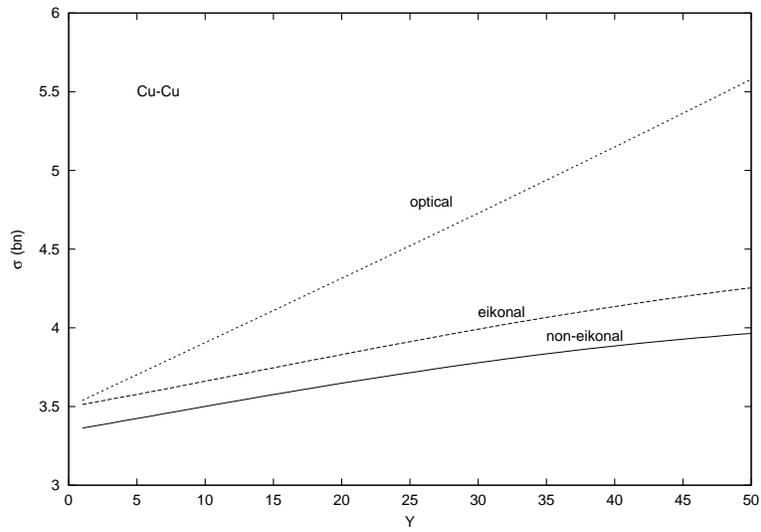}}
\caption{Total inelastic CU-Cu  cross-sections obtained with the 
non-eikonal and eikonal
boundary conditions as well as in the optical approximation as
a function of rapidity}
\label{fig6}
\end{figure}
%
\section{Inclusive cross-sections}
In the old LRFT the inclusive cross-sections are obtained by
attaching to the pomeron propagator emission verteces 
depending on the transverse momenta of the emitted particles.
They are external to the model and are to be extracted from the experimental
data. We shall not attempt to find these verteces and thus the absolute
values of the inclusive spectra. Our interest lies in rapidity correlations,
for which only inclusive spectra integrated over tranversal momenta are
important and moreover the integrated emission verteces cancel in the
expression for the correlation coefficient. So the formulas which we 
present
below refer to the inclusive cross-sections without emission verteces.
These formulas were obtained in ~\cite{CM} and we write them in their
original notation with a slight change in the overall normalization
corresponding to the standard definition of the single and double
inclusive cross-sections (see Appendix).
We stress that the approximation of zero zlope
($\alpha'=0$) is adopted. We also restrict ourselves to the simple case
of collisions of identical nuclei at a fixed impact parameter $B$.

The single inclusive cross-section is then given by an integral over all
transverse points $b$ inside the colliding nuclei
\beq
I_1(Y,y,B)=\int d^2b I_1(Y,y,B,b),
\eeq
where $Y$ is the overall rapidity of the collision, $y$ is the rapidity
of the observed particle and the inclusive cross-section $I_1(Y,y,B,b)$
at a given $b$ is  a product of two sets of fan diagrams which start
at the emission rapidity and go towards the colliding nuclei:
\beq
I_1(Y,y,B,b)\equiv\frac{d\sigma}{dyd^2b}(Y,y,B,b)=2\chi\tilde{\chi}.
\eeq
Here $\chi$ is the sum of fan diagrams going from the emission point
at rapidity $y$ to the target assumed to have rapidity zero:
\beq
\chi=\frac{AgT_A(b)e^{\mu y}}
{1+AgT_A(b)\frac{\lambda}{\mu}\left(e^{\mu y}-1\right)}
\eeq
and $\tilde{\chi}$ is the sum of fan diagrams going from the emission point
to the projectile at rapidity $Y$:
\beq
\tilde{\chi}=\chi\Big( y\to Y-y,\ T_A(b)\to T_A(B-b)\Big).
\eeq

The double inclusive cross-section for the emission at rapidity points
$y_1$ and $y_2$ contains two terms. One is just a product of two
single inclusive cross-sections
\beq
I_2^{(1)} (Y,y_1,y_2,B)=I_1(Y,y_1,B)I_1(Y,y_2,B).
\eeq
The other represents the triple pomeron and rescattering corrections.
It is expressed via the pomeron Green functions $G(y_1,y_2)$ and
$\tilde{G}(y_1,y_2)$ describing propagation of the pomeron in the
external field created by the sums of fan diagrams $\chi$ and $\tilde{\chi}$
respectively:
\beq
G(y_1,y_2)=\theta(y_1-y_2)e^{\mu(y_1-y_2)}
\Big[\frac{1+AgT_A(b)\frac{\lambda}{\mu}\left(e^{\mu y_2}-1\right)}
{1+AgT_A(b)\frac{\lambda}{\mu}\left(e^{\mu y_1}-1\right)}\Big]^2
\eeq
and
\beq
\tilde{G}(y_1,y_2)=G(y_1,y_2)\Big(y_1\to Y-y_2,\ y_2\to Y-y_1,\
T_A(b)\to T_A(B-b)\Big).
\eeq
The second part of the double inclusive is then
\[
I_2^{(2)} (Y,y_1,y_2,B)=
2\tchi(y_1)\chi(y_2)\Big(G(y_1,y_2)+
2\lambda (G\chi\tg)_{y_1y_2}\Big)\]
\beq+
8\lambda\tchi(y_1)\tchi(y_2)\Big(G\chi G^T\Big)_{y_1y_2}+
\Big(y_{1,(2)}\to Y-y_{1(2)},T_A(b)\to T_A(B-b)\Big).
\eeq
The diagrammatic representation of these contributions can be found in
~\cite{CM}. We have to stress that at $y_1=y_2$ the two contributions to
the double inclusive cross-section do not include emission of both particles at the
same rapidity illustrated in Fig. 7. This contribution is important for
correlations and should be added  as
\beq
\Delta I_2(Y,y_1,y_1)=C\tchi(y_1)\chi(y_1),
\label{add}
\eeq
where the unknown coefficient $C$ represents the result of integration of the
double emission vertex in Fig. 7 over transverse momenta of both particles
divided by the square of integrated single emission verteces.
\begin{figure}
\epsfxsize 1.3 in
\centerline{\epsfbox{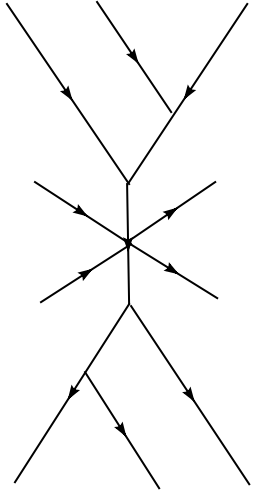}}
\caption{Diagrams with double emission at the same rapidity taken into account
by (\ref{add})}
\label{fig7}
\end{figure}

Passing to the correlation coefficient we shall restrict ourselves to
a situation when the forward window at $y_1$ and backward window
at $y_2<y_1$ are symmetrically situated relative to the center:
\beq
y_1=\frac{1}{2}(Y+\Delta y),\ \ y_2=\frac{1}{2}(Y-\Delta y),
\eeq
with $\Delta y=y_1-y_2$ being the rapidity distance between the windows.
We assume the width of both windows equal and small. Then
at fixed impact parameter $B$ the correlation coefficient is given by
\beq
\sigma_{FB}(Y,\Delta y,B)=
\frac{\sigma^{in}(B)I_2(Y,y_1,y_2,B)-I_1(Y,y_1,B)I_1(Y,y_2,B)}
{\sigma^{in}(B)I_2(Y,y_1,y_1,B)-I_1^2(Y,y_1,B)},
\eeq
where in the denominator $I_2(Y,y_1,y_1,B)$ includes the additional
contribution (\ref{add}).
In this expression everything is known except coefficient $C$ in (\ref
{add}), which remains a parameter to be extracted from the comparison with 
the
experimental data. This comparison with the RHIC data in ~\cite{expcor}
fixes $C=2.8$.

Once $C$ is fixed, the formalism allows us to find $\sigma_{FB}$ at all
values of $Y$, $\Delta y$ and $B$. In Figs. 8-10 we show our results
for Au-Au collisions at
values of $Y$ corresponding to RHIC and LHC energies and also to $Y=24$
to illustrate the general trend with the growth of energy.
In Fig. 8 and 9 we show the $\Delta y$-dependence for central and peripheral
collisions respectively. In Fig. 10 we show the $B$-dependence at
$\Delta y$=2.

As we see, the correlation coefficient $\sigma_{FB}$
grows with $Y$ towards unity
and at fixed $Y$ falls with $\Delta y$ with a slope which goes down with
$Y$ and rises with $B$. As to the behaviour of $\sigma_{FB}$
in $B$, it  practically
remains constant until the collision becomes very peripheral, when it
rapidly drops to zero. This behaviour generally agrees with
experimental observations.

Calculations for  Cu-Cu collisions show that the correlation coefficient
is practically independent of the atomic number except that the drop
for peripheral collisions occurs at correspondingly lower values of $B$. 
\begin{figure}
\epsfxsize 4 in
\centerline{\epsfbox{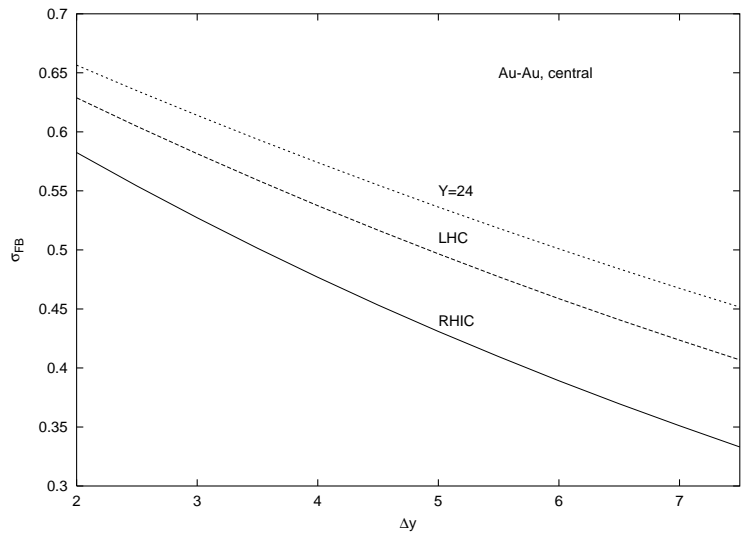}}
\caption{Correlation coefficient for central Au-Au collisions as a function
of distance between the forward and backward rapidity windows.
Curves from bottom to top correspond to energies of RHIC, LHC and
$Y=24$}
\label{Fig8}
\end{figure}
\begin{figure}
\epsfxsize 4 in
\centerline{\epsfbox{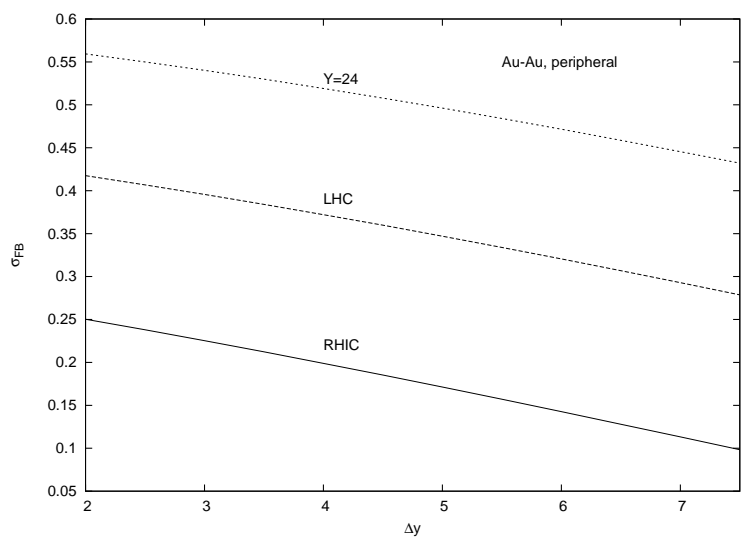}}
\caption{Same as in Fig. 8 for peripheral collisions}
\label{fig9}
\end{figure}
\begin{figure}
\epsfxsize 4 in
\centerline{\epsfbox{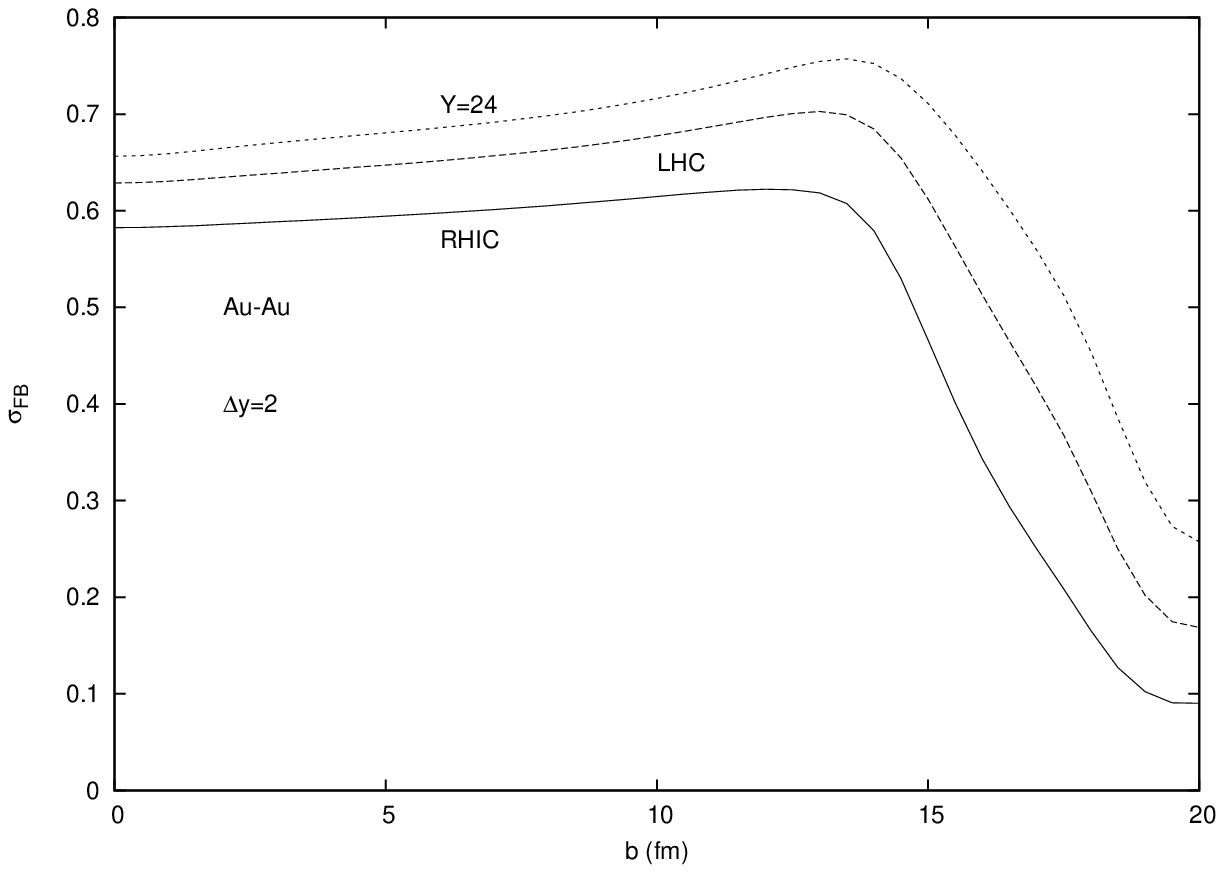}}
\caption{Correlation coefficient for  Au-Au collisions as a function
impact parameter $B$  for $\Delta y=2$.
Curves from bottom to top correspond to energies of RHIC, LHC and
$Y=24$}
\label{fig10}
\end{figure}
%
\section{Dicussion}
We have studied the nucleus-nucleus interaction in the framework of the 
old local
Reggeon Field Theory with a supercritical pomeron  having a vanishing 
slope
$\alpha'$. Loop contribution has been neglected. A new element in our 
treatment is introduction of new
boundary conditions, which substitute more common eikonal boundary
conditions to correspond to the standard Glauber expression at low 
rapidities.
The new boundary conditions are very complicated from the mathematical 
point
of view: they are non-local both in  coordinates and rapidity.

Our found iterative solutions show that the actual change in the
the action and cross-sections due to these new boundary conditions 
is not very large ($3\div 5$\% for the cross-sections). However the
structure of the solution is radically changed. Now the most complicated
region is that of low rapidities, where there exist only
a pair of asymmetric solutions and no symmetric one.
This situation is reminiscent to that in the old studies of the pure
Glauber approximation (without loops), when it was found that symmetric
solutions exist only in the limited interval
of atomic numbers  ~\cite{PTUC}.

The situation with the local pomeron is to be contrasted to that with the 
BFKL pomeron, where non-eikonal boundary conditions lower the action
at $B=0$ significantly ~\cite{BB} and one can expect a corresponding
drop in the total cross-sections.

The long-range rapidity correlations extracted from the single and
double inclusive cross-sections and the found total cross-sections
behave with overall energy, centrality and rapidity distance
in accordance with conjectures made in another approaches. 
Adjusting our only parameter to the experimental value of the correlation 
coefficient $\sigma_{FB}$ at RHIC energy, $B=0$, and $\Delta y=2$ we have 
predictions 
for $\sigma_{FB}$ at any values of energy, $B$ and $\Delta y$. 
Measurement of $\sigma_{FB}$ at these experimental conditions
will shed new light  on the validity of this simple
local supercritical pomeron approach.

\section{Acknowledgements}
The author benefited from numerous constructive discussions
with N.Armesto, S.Bondarenko, C.Pajares and Yu.Shabelsky.
The author thanks MEC of Spain for financial support under
the project FPA 2005-01963. This work was also supported 
by grants RNP 2.1.1.1112 and RFFI 06-02-16115a of Russia.

\section{Appendix. Normalization of reggeon amplitudes}
To relate the amplitudes obtained within LRFT with the physical
cross-sections it is sufficient to study the simple case $\lambda=0$
corresponding to the pure Glauber rescattering. At high energies
the relativistic forward scattering amplitude $A$ is related to the total
cross-section   by the standard relation $\sigma^{tot}={\rm Im}\, A/s$.
Passing to the LRFT theory, which lives in the purely transversal space,
one has to integrate over longitudinal variables in each loop.
This integration  gives a factor $1/(2s)$ for each loop. The total
coefficient for $n$ pomeron exchanges and thus $(n-1)$ integration loops
is  $g^{2n}s^n\cdot (2s)^{1-n}=2s(g^2/2)^n$. So factorization of the
contribution into the product of exchanges generates factor $2s$.
The cross section obtained after division by $s$ aquires factor 2.
So the correct relation between the cross-section and the amplitude in
LRFT at fixed $B$ is
\beq
 \sigma^{tot}(B)=2\ {\rm Im}\, A_{LRFT}(B)=2\Big(1-S(B)\Big).
\eeq
The elastc and inelstic cross-sections are then
\beq
 \sigma^{el}(B)=\Big(1-S(B)\Big)^2,\ \
\sigma^{in}(B)=1-S^2(B).
\eeq

With these relations in view one has to take into account that the standard
(experimental) definitions of the coupling constants $g$ and $\lambda$
are taken from the cross-sections. This implies that if $g_E$ and $\lambda_E$
are the experimental values then the parameters $g$ and $\lambda$
in the LRFT Lagrangian are determined by the relations
$
g^2_E=2g^2,\ \ g^3_e\lambda_E=2g^3\lambda
$
from which it follows
\beq
g=g_E/\sqrt{2},\ \ \lambda=\lambda_E\sqrt{2}.
\eeq
These definitions are used in (\ref{par}).


\end{document}